\documentclass[prb,twocolumn,superscriptaddress,floatfix,showkeys]{revtex4}

\usepackage{graphicx}
\usepackage{amsmath,subfigure,epsfig,psfrag}
\usepackage{comment}
\usepackage{commath}
\usepackage{graphicx}
\usepackage{amssymb}
\usepackage{xcolor}
\usepackage{soul}
\usepackage{listings}
\usepackage{rotating}
\usepackage{latexsym}
\usepackage[overload]{empheq}
\usepackage{amsfonts}
\usepackage{braket}
\usepackage{amssymb}
\usepackage{wasysym}
\usepackage{xfrac}
\usepackage{mathtools}
\usepackage{threeparttable}
\usepackage{appendix}
\makeatletter
\renewcommand*\env@matrix[1][\arraystretch]{%
  \edef\arraystretch{#1}%
  \hskip -\arraycolsep
  \let\@ifnextchar\new@ifnextchar
  \array{*\c@MaxMatrixCols c}}
\makeatother

\begin{document}

\title{Anharmonic Terms of the Potential Energy Surface: A Group Theoretical Approach}

\author{Davide Mitoli}
\affiliation{Dipartimento di Chimica, Universit\`{a} di Torino, via Giuria 5, 10125 Torino, Italy}

\author{Jefferson Maul}
\affiliation{Dipartimento di Chimica, Universit\`{a} di Torino, via Giuria 5, 10125 Torino, Italy}

\author{Alessandro Erba}
\email{alessandro.erba@unito.it}
\affiliation{Dipartimento di Chimica, Universit\`{a} di Torino, via Giuria 5, 10125 Torino, Italy}

\date{\today}

\begin{abstract}
In the framework of density functional theory (DFT) simulations of molecules and materials, anharmonic terms of the potential energy surface are commonly computed numerically, with an associated cost that rapidly increases with the size of the system. 
Recently,  an efficient approach to calculate cubic and quartic interatomic force constants in the basis of normal modes [Theor. Chem. Acc., {\bf 120}, 23 (2008)] was implemented in the \textsc{Crystal} program  [J. Chem. Theory Comput., 15, 3755-3765 (2019)]. By applying group theory, we are able to further reduce the associated computational cost, as the exploitation of point symmetry can significantly reduce the number of distinct atomically displaced nuclear configurations to be explicitly explored for energy and forces calculations. Our strategy stems from Wigner's theorem and the fact that normal modes are bases of the irreducible representations (irreps) of the point group. The proposed group theoretical approach is implemented in the \textsc{Crystal} program and its efficiency assessed on six test case systems: four molecules (methane, CH$_4$; tetrahedrane, C$_4$H$_4$; {\it cyclo}-exasulfur, S$_6$; cubane, C$_8$H$_8$), and two three-dimensional crystals  (Magnesium oxide, MgO; and a prototypical Zinc-imidazolate framework, ZIF-8). The speedup imparted by this approach is consistently very large in all high-symmetry molecular and periodic systems, peaking at 76\% for MgO.
\end{abstract}


\maketitle

\section{Introduction}

Atomic vibrational dynamics is an underlying factor to a variety of physical and chemical phenomena in molecules and materials.~\cite{BORNHUANG,THERMOSTAT_HILL} In quantum chemistry and physics, the simplest vibrational model is represented by the harmonic approximation (HA) to the Born-Oppenheimer potential energy surface (PES), where higher-than-quadratic terms are neglected and the vibration dynamics is described by a set of independent quantum harmonic oscillators.~\cite{maradudin1963} The HA has represented and still largely represents the standard approach to vibration dynamical investigations in molecules and materials because of its relatively low computational cost (as only second-order energy derivatives with respect to atomic displacements are required) and availability of robust implementations in most quantum chemistry and density functional theory (DFT) programs.~\cite{BARONI,TOGO20151,319,GONZE} We note that most solid state DFT programs implement analytic forces while the Hessian matrix is either computed analytically~\cite{giannozzi2009quantum} or from numerical finite differences of the forces computed at displaced atomic configurations.~\cite{HESSvsOPT,IntStrain}

In order to go beyond the HA and account for couplings among the normal modes of vibration, high-order terms of the PES need to be explicitly computed. 
In quantum chemistry software packages, implementations to compute higher-than-quadratic terms of the PES are scarce; moreover, explicit analytic expressions have been derived only for molecular Hartree-Fock (HF) and DFT
%When it comes to computing higher-than-quadratic terms of the PES, explicit analytic expressions have been derived only for molecular Hartree-Fock (HF) and DFT and are not commonly implemented in quantum chemistry software packages 
(see Ref. [\onlinecite{ringholm2014analytic}] and references therein for a detailed review on the evolution of analytic total energy derivatives at different orders for molecules). Common strategies for the calculation of cubic and quartic terms of the PES (i.e. third- and fourth-order total energy derivatives with respect to nuclear displacements) involve numerical differentiation making use of either just the energy or the energy and lower order analytic derivatives at a set of atomically displaced nuclear configurations. Many different numerical schemes have been proposed, each requiring a specific number of nuclear configurations to be explicitly explored.~\cite{martin1996there,burcl2003vibrational,yagi2004ab,barone2005anharmonic} We refer to Ref. [\onlinecite{Lin}] and references therein for a comprehensive review of different numerical differentiation schemes for cubic and quartic interatomic force-constants for molecular systems. The computational cost associated to each scheme is determined by two factors: the number of configurations $N_\textup{conf}$ needed and the type of calculation required at each configuration (energy only, energy and analytic forces, etc.). The former factor accounts for most of the computational cost as computing the energy through the self-consistent field (SCF) procedure at a new nuclear configuration proves much more expensive than analytically evaluating the forces after the SCF is completed at an already explored geometry. 

In a solid state context, much effort has gone into the implementation of schemes for the description of cubic terms of the PES (i.e. those relevant to the computation of the lattice thermal conductivity),~\cite{togo2015-3py,togo2015-py,plata2017,skelton2014,whalley2016,linnera2017} with fewer attempts to include up to quartic terms.~\cite{PhysRevLett.113.185501,prentice2017,souvatzis2008,errea2014,parlinski2018} Let us note that, for solids, the HA limits the description of the lattice dynamics even more than it does for molecular systems because of the corresponding missing dependence of thermodynamic properties on volume. However, this class of limitations can be largely overcome by computing harmonic phonons at different lattice volumes within the so-called quasi-harmonic approximation (QHA).~\cite{QHA,Baroni01012010} Some of the authors of this paper have recently developed a module of the \textsc{Crystal} program for the calculation of quasi-harmonic thermal properties of materials (from thermal expansion to thermo-elasticity).\cite{Grun_MIO,Thermal_MgCaO,CORUNDUM,FORSTERITE,LIF_PRL,UREA_QHA,THERMOELAST_MIO,ryder2019quasi,CU2O_ANTTI,THERMOELA_ORGSEMI,maul2020thermoelasticity}

A numerically robust and computationally efficient finite difference scheme (namely, EGH), based on a Taylor's expansion of the PES in the basis of the normal modes, has been proposed in 2008 for molecular systems by Lin {\it et al.}, which requires a minimal set of nuclear configurations to be explored in the definition of a 2M4T or 3M4T representation of the PES.~\cite{Lin} Such scheme (based on the analysis of the relative importance of different types of cubic and quartic terms) has recently been extended to solids by some of the present authors~\cite{PARTI_ANHARM} and implemented in the \textsc{Crystal} program,~\cite{CRYSTAL17PAP,PAPCRYSTAL23} along with the vibrational self-consistent field (VSCF) and vibrational configuration interaction (VCI) methods for computation of anharmonic vibrational states.~\cite{PARTII_ANHARM,maul2019elucidating,schireman2022anharmonic}

In this paper, we illustrate how group theoretical arguments can be used to drastically reduce the number of configurations $N_\textup{conf}$ needed to achieve a quartic representation of the PES on both molecules and materials belonging to high point symmetry groups. An algorithm is presented, as implemented in a developmental version of the \textsc{Crystal23} program, whose efficiency is documented by numerical tests performed on selected molecules and crystalline materials.

\section{Formal Aspects}
\label{sec:form}

\subsection{The Truncation of the PES}

By computing, mass-weighting and diagonalizing the Hessian matrix of either a molecular system with $N$ atoms or a crystal with $N$ atoms per cell, normal modes $Q_i$ and associated harmonic vibration frequencies $\omega_i$ are obtained, with $i=1,\dots, M$, where $M=3N -6(5)$ for molecules and $M=3N -3$ for solids at the $\Gamma$ point. Within the Born-Oppenheimer approximation, vibrational states are determined by solving the nuclear Schr\"odinger equation, which, in terms of normal coordinates, reads:
\begin{equation}
\label{eq:scrod}
\hat{H} \Psi_s({\bf Q}) = E_s \Psi_s({\bf Q}) \; ,
\end{equation}  
where $\Psi_s({\bf Q})$ is the vibrational wavefunction of the $s$-th vibrational state and $E_s$ the corresponding energy.
By setting the rotational angular momentum to zero and by neglecting rotational coupling effects, the Hamiltonian operator in Eq. (\ref{eq:scrod}) can be written as:
\begin{equation}
\label{eq:hamil}
\hat{H} = \sum_{i=1}^M -\frac{1}{2} \frac{\partial^2}{\partial Q_i^2} + \hat{V}(Q_1, \dots, Q_M) \; .
\end{equation}
The Born-Oppenheimer PES can be expanded in a Taylor's series centered at the equilibrium nuclear configuration in the basis of such mass-weighted normal coordinates:  
\begin{eqnarray}
\label{eq:pes}
\hat{V}(Q_1, \dots, Q_M) &=& \frac{1}{2}\sum_{i=1}^M \omega_i^2 Q_i^2  \nonumber \\
&+& \frac{1}{3!} \sum_{i,j,k=1}^M \eta_{ijk} Q_iQ_jQ_k + \nonumber \\
&+&  \frac{1}{4!} \sum_{i,j,k,l=1}^M \eta_{ijkl} Q_iQ_jQ_kQ_l  + \cdots \; , \;\;\;
\end{eqnarray}
where $\eta_{ijk}$ and $\eta_{ijkl}$ are cubic and quartic force constants, respectively:
\begin{eqnarray}
\label{eq:fconst}
\eta_{ijk} &=& \left( \frac{\partial^3 E}{\partial Q_i \partial Q_j \partial Q_k}\right)_\textup{eq} \\
\label{eq:fconst2}
\eta_{ijkl} &=& \left( \frac{\partial^4 E}{\partial Q_i \partial Q_j \partial Q_k \partial Q_l}\right)_\textup{eq} \; .
\end{eqnarray}
These are high-order total energy derivatives with respect to collective normal coordinates, evaluated at the equilibrium nuclear configuration. The PES expansion in Eq. (\ref{eq:pes}) needs to be truncated so as to include only those terms contributing significantly to the description of the vibrational states of the system.

In molecular anharmonic calculations, it is a common practice to truncate it after the fourth-order as in most cases neglected higher-than-quartic terms would produce little corrections to the vibrational states (note that for strongly anharmonic systems such as water this may not be the case).~\cite{Lin}  Here, we follow the same strategy and thus we consider terms up to fourth-order in the PES (namely, we use a quartic, 4T, representation of the potential). Within a 4T representation, the PES can be further truncated by considering only those force constants involving a maximum of $n$ distinct modes (namely, a $n$M representation of the potential). By combining the two truncation strategies introduced above, a 1M4T representation of the PES would require the evaluation of the force constants below:
\begin{equation}
\label{eq:1M4T}
\eta_{iii} , \; \eta_{iiii} \quad \forall \; i=1,\dots, M \; . 
\end{equation}
This representation of the PES neglects two-mode couplings and almost always results in a wrong description of the vibrational states. A popular representation of the potential is the 2M4T one, which includes all two-mode coupling force constants while neglecting three- and four-mode terms:~\cite{Lin} 
\begin{eqnarray}
\label{eq:2M4T}
\eta_{iii} , \; \eta_{iiii} \quad &\forall& \; i=1,\dots, M  \nonumber \\
\eta_{ijj} , \; \eta_{iij}  , \; \eta_{iiij}  , \; \eta_{ijjj}  , \; \eta_{iijj} \quad &\forall& \; i<j  =1,\dots, M \;.  \; \;
\end{eqnarray}
This is the representation of the PES we work with here.

\begin{figure}[h!!]
\centering
\includegraphics[width=8.6cm]{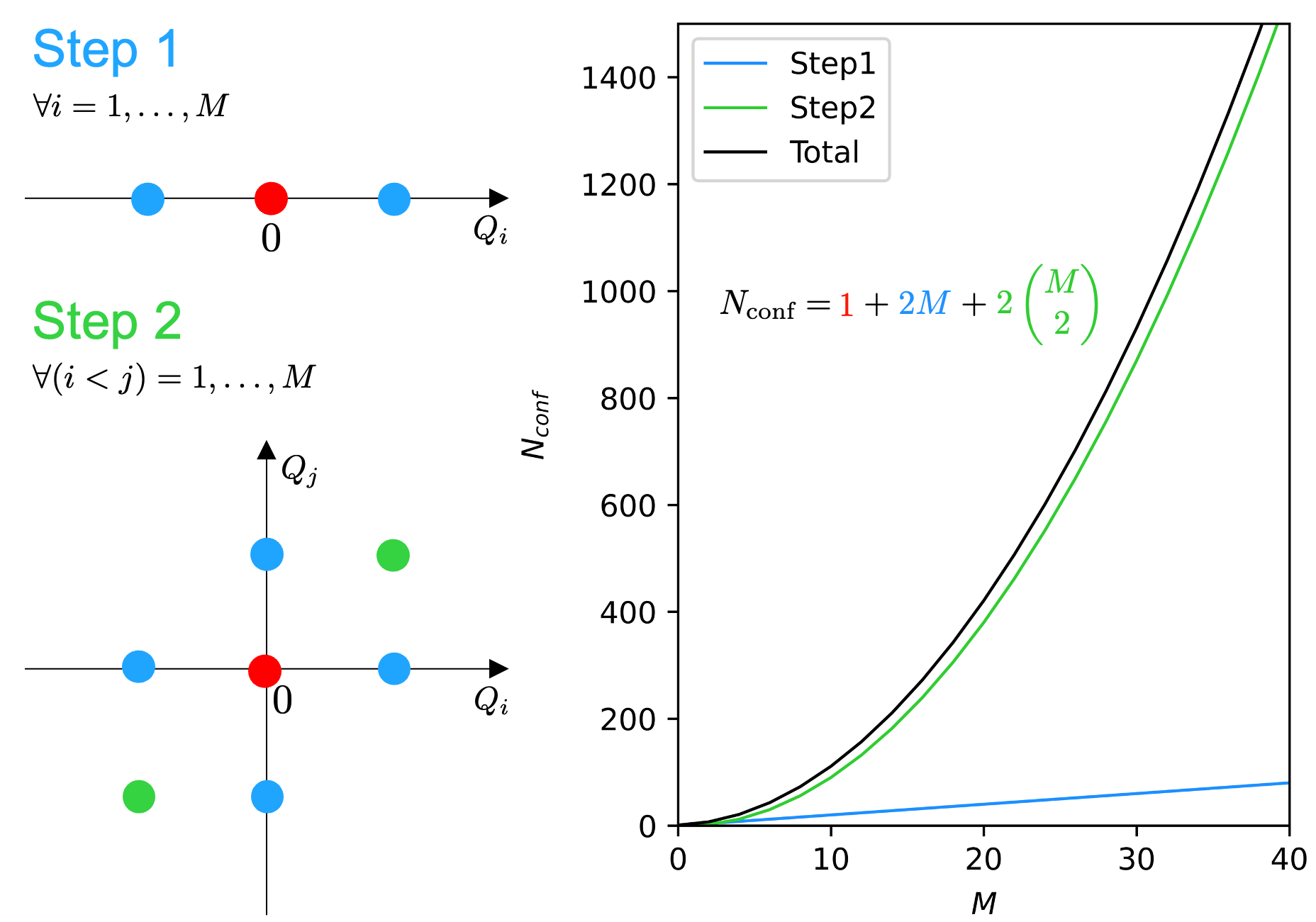}
\caption{(Left) Schematic representation of the displaced nuclear configurations required by the EGH scheme for a 2M4T description of the PES; (Right) Dependence of the number of nuclear configurations $N_\textup{conf}$ needed for the EGH scheme as a function of the number of normal modes $M$.}
\label{fig:egh}
\end{figure}

\subsection{The EGH Scheme for Cubic and Quartic Terms of the PES}
\label{sec:egh}

We start by briefly recalling the fundamentals of the EGH finite-difference scheme, as originally proposed by Lin {\it et al.}~\cite{Lin} We refer to Figure \ref{fig:egh} for a schematic representation. The zeroth step consists in the evaluation of the Hessian at the equilibrium configuration to obtain the harmonic normal modes and frequencies. Then, as a first step, for each normal coordinate $Q_i$, two nuclear configurations are explored towards positive and negative atomic displacements relative to the equilibrium configuration (blue circles in Figure \ref{fig:egh}). Both the total energy $E$ and the analytical gradients $G$ are computed at these configurations. The corresponding 1M terms of the PES of each mode $Q_i$ can be obtained from:
\begin{eqnarray}
\label{eq:scheme3step11}
\eta_{iii}  &=& \dfrac{1}{s_i^2}\left( G_{-1}^{i} + G_{+1}^{i} \right) \\
\label{eq:scheme3step12}
\eta_{iiii} &=& \dfrac{3}{s_i^3}\left( G_{+1}^{i} -2s_i\omega_{i} - G_{-1}^{i} \right) \; ,
\end{eqnarray}
where $s_i = h/\sqrt{\omega_i}$ is an adaptive step (see Ref. [\onlinecite{PARTI_ANHARM}] for more details on the definition of the step size $h$) and $G^i_a$ is the gradient with respect to $Q_i$ computed at a nuclear configuration displaced by $a \cdot s_i \cdot Q_i$ from the equilibrium one.

As a second step, for each pair of normal modes $(Q_i, Q_j)$ so that $i<j$, two nuclear configurations are explored with positive and negative atomic displacements from the equilibrium configuration along both modes at the same time (green circles in Figure \ref{fig:egh}). For a 2M4T representation of the PES, just the total energy $E$ is required at these configurations. The corresponding 2M terms of the PES for each pair of modes are obtained from:  
\begin{eqnarray}
\label{eq:scheme3step21}
\eta_{iij}  &=& \dfrac{1}{s_i^2}( G_{-1,0}^{j} + G_{1,0}^{j} ) \\ 
\label{eq:scheme3step22}
\eta_{iiij} &=& \dfrac{3}{s_i^3}( G_{1,0}^{j} - G_{-1,0}^{j} ) \\
\label{eq:scheme3step23}
\eta_{iijj} &=& -\dfrac{1}{2s_i^2s_j^2}( 8E_{0,0} - 4 E_{-1,-1} -4E_{1,1} + \nonumber \\
&&-s_jG_{0,-1}^{j} + s_jG_{0,1}^{j} - s_iG_{-1,0}^{i} + s_iG_{1,0}^{i} + \nonumber\\
&&-4s_jG_{-1,0}^{j}+4s_jG_{1,0}^{j} - 4s_iG_{0,-1}^{i}+4s_iG_{0,1}^{i} + \nonumber\\
&&+2s_i^2\omega_{i}+ 2s_j^2\omega_{j} ) \; ,
\end{eqnarray}
where $E_{a,b}$ and where $G_{a,b}^i$ are the total energy and the gradient with respect to $Q_i$ computed at a nuclear configuration displaced by $a \cdot s_i \cdot Q_i + b \cdot  s_j \cdot Q_j$ from the equilibrium one, respectively. Analogously, $G_{a,b}^j$ is the gradient with respect to $Q_j$ computed at the same nuclear configuration. For a system with $M$ normal modes, the total number of nuclear configurations to be explored in the definition of the 2M4T PES with this scheme is therefore given by:
\begin{equation}
\label{eq:count3}
N_\textup{conf} = 1 + 2M + 2 \binom{M}{2} \; .
\end{equation}
Figure \ref{fig:egh} shows the dependence of $N_\textup{conf}$ on $M$.

\subsection{From Wigner's Theorem to Symmetry Relations among Terms of the Anharmonic PES}
\label{subsec:symm}

The point symmetry group $\mathcal{P}$ of a system (molecular or crystalline) is the set of those point symmetry operators $\hat{R}_r$ (with $r = 1,\dots, \vert \mathcal{P}\vert$, where $\vert \mathcal{P}\vert$ is the so-called order of the group) with respect to which the system is invariant. This is expressed by Wigner's theorem that states that each point symmetry operator of the group must commute with the Hamiltonian operator:~\cite{wigner1931gruppentheorie,ZicErb2,ZicErb}
\begin{equation}
\label{eq:wigner}
\left[ \hat{H}, \hat{R}_r \right] = 0 \qquad \forall r = 1, \dots, \vert \mathcal{P}\vert \; .
\end{equation}
We note that if the Hamiltonian operator consists of a sum of terms, the condition above must be satisfied by each term individually. In group theory, a finite number $N_\textup{irrep}$ of irreducible representations (irreps) can be associated to a point group. Each irrep $\Gamma_\alpha$ (with $\alpha=1,\dots,N_\textup{irrep}$) has a given dimensionality $n_\alpha$ (for standard molecular or crystalline point groups $n_\alpha=1,2,3$) and is fully characterized by the corresponding $n_\alpha\times n_\alpha$ representation matrices $D^\alpha(\hat{R}_r)$ associated to each point symmetry operator. The characters $\chi^\alpha(\hat{R}_r)$ of a representation are simply defined as the trace of these matrices: $\chi^\alpha(\hat{R}_r) = \textup{Tr}[D^\alpha(\hat{R}_r)]$. Each irrep $\Gamma_\alpha$ is defined by $n_\alpha$ basis functions, corresponding to each row of the representation matrices. Let us introduce so-called projector operators associated to each irrep of the group:
\begin{equation}
\label{eq:proj}
\hat{P}^\alpha  = \frac{1}{\vert \mathcal{P}\vert} \sum_{r=1}^{\vert \mathcal{P}\vert}  \chi^\alpha(\hat{R}_r)^\ast  \hat{R}_r  \; .
\end{equation}
This operator is such to act on any function and ``extract'' its components of $\alpha$ type.

Normal modes are bases of the irreps of the point symmetry group of the system. Different modes can belong to the same irrep so that overall the manifold of all the harmonic modes of a system can be expressed as a direct sum of irreps as:
\begin{equation}
\Gamma_\textup{HA} = m_1\Gamma_1 \oplus  \cdots \oplus m_\alpha\Gamma_\alpha \oplus \cdots \oplus  m_{N_\textup{irrep}}\Gamma_{N_\textup{irrep}} \, ,
\end{equation}
where $m_\alpha$ is the multiplicity of irrep $\alpha$ in the manifold. The group theoretical approach that we illustrate requires normal modes to be explicitly labeled according to their symmetry properties. Therefore, we introduce the following extended notation for each normal mode:
\begin{equation}
\label{eq:not}
Q_i \rightarrow Q_{\alpha ul} \equiv \ket{\alpha u l} \; ,
\end{equation}
where each normal mode is labeled by the irrep $\alpha$ it belongs to, an index $u = 1,\dots,n_{\alpha}$ identifying the row of the irrep it is associated to, and an index $l = 1,\dots,m_{\alpha}$ marking the occurrence of the irrep $\Gamma_\alpha$ the mode refers to. From now on, we refer to the $n_\alpha$ basis functions of each occurrence of each irrep as a {\it set}. Moreover, with the notation introduced in Eq. (\ref{eq:not}) we adopt a ket notation with the aim of highlighting these labels more prominently, to be extensively used in what follows.

To make the notation introduced with Eq. (\ref{eq:not}) more clear, we shall analyse how it works for the simple molecule of methane in Table \ref{tab:methane}. Methane, CH$_4$, is a non-linear molecule with $N=5$ atoms and $M=9$ normal modes (excluding pure translations and rotations) belonging to the T$_\textup{d}$ point symmetry group. Modes 1, 2 and 3 belong to the first occurrence of the 3D irrep $F_2$, form the first {\it set}, and are degenerate (i.e. they have the same harmonic vibration frequency). Modes 4 and 5 belong to the first occurrence of the 2D irrep $E$, form the second {\it set}, and are degenerate. Mode 6 belongs to the 1D total-symmetric irrep $A_1$ and forms the third {\it set} on its own. Finally, modes 7, 8 and 9 belong to the second occurrence of the 3D irrep $F_2$, form the fourth {\it set}, and are degenerate.

\begin{table}[b]
\centering
\caption{Symmetry features of the normal modes of methane (the six pure translations and rotations are excluded).}
\label{tab:methane}
\begin{tabular} {ccccc}
    \hline
    \hline
    Mode & Irrep $\alpha$ & Row $u$ & Occurrence $l$ & {\it Set}\\
    \hline
    1 & $F_2$ &1& 1 & 1\\
    2 & $F_2$ &2& 1 & 1\\
    3 & $F_2$ &3& 1 & 1\\
    4 & $E$ &1& 1 & 2\\
    5 & $E$ &2& 1 & 2\\
    6 & $A_1$ &1& 1 & 3\\
    7 & $F_2$ &1& 2& 4\\
    8 & $F_2$ &2& 2& 4\\
    9 & $F_2$ &3& 2& 4\\
   \hline
   \hline
\end{tabular}
\end{table}

Let us recall how, from group theory, basis functions of irreps transform upon application of a point symmetry operator:
\begin{equation} 
\label{operator-application}
\hat{R}_r\ket{\alpha ul} = \sum_{u'=1}^{n_{\alpha}} D_{u'u}^{\alpha}(\hat{R}_r)\ket{\alpha u'l} \; ,
\end{equation}
that is, when acted upon by a symmetry operator, each normal mode $\ket{\alpha ul}$ is transformed into a linear combination of the basis functions of the {\it set} it belongs to, with coefficients given by the elements of the corresponding representation matrix. Given that cubic and quartic terms of the PES involve products of three or four normal modes, it will prove useful to what follows to show how a symmetry operator acts on a direct product of such functions, through its linearity property:
\begin{equation} 
\label{operator-application2}
\hat{R}_r \left( \ket{\alpha ul} \otimes \cdots \otimes \ket{\omega zs} \right) =\hat{R}_r  \ket{\alpha ul} \otimes \cdots \otimes \hat{R}_r \ket{\omega zs} \; .
\end{equation}
Wigner's theorem, as introduced in Eq. (\ref{eq:wigner}), proves key to an effective exploitation of point-symmetry to reduce the cost of the evaluation of an anharmonic PES such as the 2M4T one, that is to reduce the number of nuclear configurations $N_\textup{conf}$ to be explored to compute all of the interatomic force constants in Eq. (\ref{eq:2M4T}). Each additive term of the Hamiltonian (\ref{eq:hamil}) and thus of the potential (\ref{eq:pes}) must be invariant to any symmetry operator of the group. Taking into account that cubic and quartic terms of the PES are nothing but products of three or four normal modes, this can be formally expressed as:
$$
\ket{\alpha u l} \otimes \cdots \otimes \ket{\omega z s} \equiv 
\hat{R}_r ( \ket{\alpha u l} \otimes \cdots \otimes \ket{\omega z s})  \; .
$$ 
Because the invariance of each term of the PES has to be satisfied for every symmetry operator, the expression above can also be written as:
\begin{equation}
\label{eq:lll}
\ket{\alpha u l} \otimes \cdots \otimes \ket{\omega z s} \equiv 
\frac{1}{\vert \mathcal{P}\vert} \sum_{r=1}^{\vert \mathcal{P}\vert} \hat{R}_r ( \ket{\alpha u l} \otimes \cdots \otimes \ket{\omega z s})  \; .
\end{equation}
We note that by comparison with Eq. (\ref{eq:proj}) and by recalling that all characters of the total-symmetric irrep A$_1$ are 1 (i.e. $\chi^{\textup{A}_1}(\hat{R}_r) = 1 \;\; \forall r=1,\dots,\vert \mathcal{P}\vert$), Eq. (\ref{eq:lll}) can be written as:
\begin{equation}
\label{eq:lll2}
\ket{\alpha u l} \otimes \cdots \otimes \ket{\omega z s} \equiv 
\hat{P}^{\textup{A}_1} ( \ket{\alpha u l} \otimes \cdots \otimes \ket{\omega z s})  \; .
\end{equation}
Now, by casting Eq. (\ref{operator-application2}) into Eq. (\ref{eq:lll}) and by use of property (\ref{operator-application}), the invariance condition for any general term of the PES becomes:
\begin{widetext}
\begin{eqnarray}
\label{eq:general_symm_rel}
\ket{\alpha u l} \otimes \cdots \otimes \ket{\omega z s} & \equiv&
    \frac{1}{|\mathcal{P}|} \sum_{r=1}^{|\mathcal{P}|} \hat{R}_r   (\ket{\alpha u l} \otimes \cdots 
    \otimes \ket{\omega z s} \nonumber \\ 
 &=&   \frac{1}{|\mathcal{P}|} \sum_{r=1}^{|\mathcal{P}|}    \hat{R}_r  \ket{\alpha ul} \otimes \cdots \otimes \hat{R}_r \ket{\omega zs}        \nonumber \\
   &=& \frac{1}{|\mathcal{P}|}\sum_{r=1}^{|\mathcal{P}|} 
    \sum_{u'=1}^{n_\alpha} \cdots
    \sum_{z'=1}^{n_\omega} D_{u'u}^\alpha (\hat{R}_r)\cdots
    D_{z'z}^\omega(\hat{R}_r)\ket{\alpha u' l} \otimes \cdots \otimes 
    \ket{\omega z' s} \nonumber \\
    &=& \sum_{u'=1}^{n_\alpha}\cdots\sum_{z'=1}^{n_\omega}C_{u\cdots z,
    u'\cdots z'}\ket{\alpha u' l} \otimes \cdots \otimes \ket{\omega z' s} \; ,
\end{eqnarray}
\end{widetext}
where we have introduced coefficients $C_{u\cdots z, u' \cdots z'}$ defined as:
\begin{equation}
\label{eq:coeff}
    C_{u\cdots z, u' \cdots z'} =
    \frac{1}{|\mathcal{P}|}\sum_{r=1}^{|\mathcal{P}|}D_{u'u}^\alpha(\hat{R}_r)
    \cdots D_{z'z}^\omega(\hat{R}_r) \; .
\end{equation}
The invariance condition as explicitly worked out in Eq. (\ref{eq:general_symm_rel}) constitutes our working expression for an effective symmetry analysis of the anharmonic PES. Two main scenarios can be met: 
\begin{enumerate}
\item If all coefficients are null on the right-hand side (rhs) then the term of the PES on the left-hand side (lhs) must be null;
\item If some coefficients are not null on the rhs then a symmetry relation is determined among the PES term on the lhs and those appearing on the rhs that must hold true. Let us note that those terms of the PES on the rhs involve normal modes belonging to the same {\it sets} of the modes of the PES term on the lhs. We shall call these connected terms {\it relative terms} below. We stress that a term of the PES can be related by symmetry only to those that constitute its {\it relative terms} as identified by Eq. (\ref{eq:general_symm_rel}).
\end{enumerate}
To summarize, use of Eq. (\ref{eq:general_symm_rel}) allows performing a preliminary symmetry analysis of the anharmonic terms of the PES and determining whether specific terms must be null by symmetry (and thus do not need to be explicitly computed) or whether specific symmetry relations must be satisfied among subsets of terms of the PES (i.e. among {\it relative terms}). Let us note that these symmetry relations are such that not all {\it relative terms} are independent. We will discuss below how a minimal number of terms of the PES to be explicitly computed can be identified, which then allows all others to be obtained by exploitation of such symmetry relations.

\subsection{On the Use of the Symmetry Relations among Terms of the Anharmonic PES}
\label{subsec:use}

In this Section, we discuss how Eq. (\ref{eq:general_symm_rel}) can be used to effectively reduce the number of nuclear configurations $N_\textup{conf}$ needed for the evaluation of all the anharmonic interatomic force constants in Eq. (\ref{eq:2M4T}). To do this, we need to inspect Eq. (\ref{eq:general_symm_rel}) more closely. The term on the lhs can be either cubic or quartic; its {\it relative terms} on the rhs will be cubic or quartic, respectively. The number of {\it relative terms} $n_\textup{rt}$ depends on the dimensionality of the irreps involved in the term on the lhs, and is simply given by $n_\textup{rt} = n_\alpha \times \cdots \times n_\omega$. 
%Depending on the number of distinct {\it sets} of modes involved in the term on the lhs, not all $n_\textup{rt}$ will be distinct
Thus, to each term of the PES a group of {\it relative terms} can be associated that we label simply $\ket{t}$ (with $t=1,\dots,n_\textup{rt}$) with a shorthand notation. To make this more evident, we introduce the following exemplification where we consider as term on the lhs the two-mode cubic term $\ket{\alpha 1 l} \otimes \ket{\alpha 1 l} \otimes \ket{\beta 1 m}$ with $n_\alpha = 2$ and $n_\beta = 1$. The corresponding {\it relative terms} would be:
\begin{eqnarray}
&\ket{\alpha 1 l} \otimes \ket{\alpha 1 l} \otimes \ket{\beta 1 m} \to \ket{1} &\nonumber \\
& \ket{\alpha 1 l} \otimes \ket{\alpha 2 l} \otimes \ket{\beta 1 m} \to \ket{2}& \nonumber \\
& \ket{\alpha 2 l} \otimes \ket{\alpha 1 l} \otimes \ket{\beta 1 m} \to \ket{3} & \nonumber \\
&\ket{\alpha 2 l} \otimes \ket{\alpha 2 l} \otimes \ket{\beta 1 m} \to \ket{4}& \nonumber \; .
\end{eqnarray}  
By use of this simplified notation, the invariance condition of Eq. (\ref{eq:general_symm_rel}) for the first term of a group of {\it relative terms} can be written in a more compact fashion as:
\begin{equation}
\label{eq:cond1}
\ket{1} = C_{1,1} \ket{1} + C_{1,2} \ket{2} + \dots + C_{1,n_\textup{rt}} \ket{n_\textup{rt}} \; ,
\end{equation}
where the coefficients $C_{t,t^\prime}$ are those introduced in Eq. (\ref{eq:coeff}), as expressed in the new shorthand notation. The action of the invariance condition of Eq. (\ref{eq:general_symm_rel}) on the second term of the group of {\it relative terms} would lead to a symmetry relation of the form:
\begin{equation}
\ket{2} = C_{2,1} \ket{1} + C_{2,2} \ket{2} + \dots + C_{2,n_\textup{rt}} \ket{n_\textup{rt}} \; ,
\end{equation}
with different coefficients with respect to those of Eq. (\ref{eq:cond1}) but associated to the same terms of the PES (i.e. those belonging to the selected group of {\it relative terms}). These symmetry relations can be derived for each term of the PES among {\it relative terms} to form a linear system of $n_\textup{rt}$ equations of $n_\textup{rt}$ variables:
  \begin{alignat}{4}[left = \empheqlbrace]
    \ket{1} &= C_{1,1} \ket{1} + C_{1,2} \ket{2} + \dots + C_{1,n_\textup{rt}}\ket{n_\textup{rt}}
    \nonumber \\
    \ket{2} &= C_{2,1} \ket{1} + C_{2,2} \ket{2} + \dots + C_{2,n_\textup{rt}}\ket{n_\textup{rt}}
    \nonumber \\
            &\vdotswithin{\dots} \phantom{0} \nonumber\\
    \ket{n_\textup{rt}} &= C_{n_\textup{rt},1} \ket{1} + C_{n_\textup{rt},2} \ket{2} + \dots + C_{n_\textup{rt},n_\textup{rt}} \ket{n_\textup{rt}} \; .
\end{alignat}
The system above can be made homogeneous as:
\begin{alignat}{4}[left = \empheqlbrace]
\label{eq:system}
    C^\prime_{1,1} \ket{1} + C^\prime_{1,2} \ket{2} + \dots + C^\prime_{1,n_\textup{rt}} \ket{n_\textup{rt}} 
    &= 0 \nonumber \\
    C^\prime_{2,1} \ket{1} + C^\prime_{2,2} \ket{2} + \dots + C^\prime_{2,n_\textup{rt}} \ket{n_\textup{rt}} 
    &= 0 \nonumber \\
    &\vdotswithin{\dots} \phantom{0} \nonumber\\
    C^\prime_{n_\textup{rt},1} \ket{1} + C^\prime_{n_\textup{rt},2} \ket{2} + \dots + C^\prime_{n_\textup{rt},n_\textup{rt}} \ket{n_\textup{rt}} &= 0 
\end{alignat}
where the primed coefficients $C^\prime_{t,t^\prime}$ are related to the unprimed ones as:
\begin{equation}
\label{eq:c'}
    C^\prime_{t,t^\prime}=
    \begin{cases}
        \;\;\;\;1 & \text{if} \;t=t^\prime\\
        \frac{C_{t,t^\prime}}{C_{t,t} -1} & \text{if}\; t\neq t^\prime
    \end{cases}
\end{equation}
In matrix notation, the linear system of Eq. (\ref{eq:system}) can be written as:
\begin{equation}
\label{eq:system2}
\bf{C}^\prime {\bf t} = {\bf 0} \; ,
\end{equation}
where $\bf{C}^\prime$ is the $n_\textup{rt}\times n_\textup{rt}$ matrix of the $C^\prime_{t,t^\prime}$ coefficients and ${\bf t}$ is the vector of the $n_\textup{rt}$ unknown {\it relative terms}. The key point to the whole symmetry analysis of the anharmonic PES we introduce is to look for non-trivial solutions of these linear systems. For each group of {\it relative terms}, we aim at identifying the minimal set of $n_\textup{ec}$ terms of the PES to be explicitly computed via the EGH finite difference numerical approach described in Section \ref{sec:egh} that allows for the system of Eq. (\ref{eq:system2}) to be solved and thus for the other $n_\textup{s} = n_\textup{rt} - n_\textup{ec}$ terms to be obtained by symmetry. By explicitly computing $n_\textup{ec}$ terms, the number of unknown variables reduces to $n_\textup{s}$ and thus we are left with a system of $n_\textup{rt}$ equations and $n_\textup{s}$ variables:
\begin{alignat}{4}[left = \empheqlbrace]
\label{eq:inhomogeneous}
    \sum_{t_s}^{n_\textup{s}} C'_{1,t_s} \ket{t_s} &= -\sum_{t_\textup{ec}}^{n_\textup{ec}} C'_{1,t_\textup{ec}}\ket{t_\textup{ec}} \nonumber \\
    \sum_{t_s}^{n_\textup{s}} C'_{2,t_s} \ket{t_s} &= -\sum_{t_\textup{ec}}^{n_\textup{ec}} C'_{2,t_\textup{ec}}\ket{t_\textup{ec}} \nonumber \\
    &\vdotswithin{\dots} \phantom{0} \nonumber\\
    \sum_{t_s}^{n_\textup{s}} C'_{n_\textup{rt},t_s} \ket{t_s} &=  -\sum_{t_\textup{ec}}^{n_\textup{ec}} C'_{n_\textup{rt},t_\textup{ec}}\ket{t_\textup{ec}} \nonumber \\
\end{alignat}
It is important to mention that a system like this is overdetermined. A smaller system can be obtained by removing $n_\textup{ec}$ equations from the previous one so as to get a reduced square matrix of coefficients on the lhs:
\begin{equation}
\label{eq:system3}
\bf{C}^\prime_{\bf r} {\bf t}_{\bf s} = {\bf t}_{\bf ec} \; ,
\end{equation}
where $\bf{C}^\prime_{\bf r}$ is a $n_\textup{s}\times n_\textup{s}$ square matrix, ${\bf t}_{\bf s}$ is a vector whose elements are the force constants to be determined by symmetry (i.e. by solving the system of equations), and ${\bf t}_{\bf ec}$ is a vector obtained from the explicitly computed terms, with elements being the rhs of the equations in (\ref{eq:inhomogeneous}). Now, the linear system of equations (\ref{eq:system3}) can be solved if the following condition is satisfied:
\begin{equation}
\label{eq:condition}
\det {\bf C}^\prime_{\bf r} \neq 0 \; .
\end{equation}
Once this condition is met, it is trivial to obtain by symmetry the $n_\textup{s}$ terms of the PES just by inverting the $\bf{C}^\prime_{\bf r}$ matrix as:
\begin{equation}
\label{eq:solution}
{\bf t}_{\bf s} = \left[{\bf C}^\prime_{\bf r}\right]^{-1} {\bf t}_{\bf ec} \; .
\end{equation}
Condition (\ref{eq:condition}) is the one we use to devise an algorithm that performs a preliminary symmetry analysis to identify the minimal set of terms of the PES to be explicitly computed via the EGH numerical scheme that allows for the whole set of terms of a 2M4T PES to be computed.

\subsubsection{The Algorithm}

This is how the algorithm we have devised works. For each set of {\it relative terms} of the PES, the invariance condition (\ref{eq:general_symm_rel}) is applied to each term of the set and the linear system of equations (\ref{eq:system2}) is built. In order to identify the minimal set of terms among them that need to be explicitly computed via the EGH scheme to make the linear system solvable and thus obtain the remaining terms by symmetry, we exploit condition (\ref{eq:condition}) and proceed as discussed below:
\begin{enumerate}
\item From the invariance condition (\ref{eq:general_symm_rel}) we determine what terms must be null by symmetry. Let us label the number of such terms $n_0$. This leaves us with $\tilde{n}_\textup{rt} = n_\textup{rt} -n_0$ non vanishing {\it relative terms}. The $n_0$ null terms, along with the corresponding equations, can be safely eliminated from (\ref{eq:system2}), so that we are left with a linear system of $\tilde{n}_\textup{rt}$ variables and $\tilde{n}_\textup{rt}$ equations.
\item By analysing the {\it relative terms}, 3M or 4M ones are identified (i.e. those terms involving three or four distinct modes). Let the number of such terms be $n_{34}$. It is important to correctly identify these terms because the EGH scheme, as discussed in Section \ref{sec:egh} and implemented in the \textsc{Crystal} program, does not allow for their explicit calculation.
\item The process starts by checking whether the system can be solved by explicitly computing only one term (i.e. by setting $n_\textup{ec} = 1$). This involves an iterative procedure where at each iteration one of the $\tilde{n}_\textup{rt} - n_{34}$ variables is selected and moved to the rhs of (\ref{eq:inhomogeneous}). In order to reduce the overdetermined system to the form (\ref{eq:system3}), one of the $\tilde{n}_\textup{rt}$ equations must be removed. This is done in turn with an iterative procedure. For each combination in the iterative process, condition (\ref{eq:condition}) is checked. If the condition is satisfied the process stops otherwise it keeps going.  
\item If condition (\ref{eq:condition}) was never satisfied at the previous step, then the process checks whether the system can be solved by explicitly computing only two terms (i.e. by setting $n_\textup{ec} = 2$). This involves an iterative procedure where at each iteration two of the $\tilde{n}_\textup{rt} - n_{34}$ variables are selected and moved to the rhs of (\ref{eq:inhomogeneous}). In order to reduce the overdetermined system to the form (\ref{eq:system3}), two of the $\tilde{n}_\textup{rt}$ equations must be removed. This is done in turn with an iterative procedure by exploring all possible pairs of equations. For each combination in the iterative process, condition (\ref{eq:condition}) is checked. If the condition is satisfied the process stops otherwise it keeps going.  
\item If condition (\ref{eq:condition}) was not satisfied at steps 3 and 4 above, the process goes on by setting $n_\textup{ec} = 3,4,\dots$ until the condition is met.
\end{enumerate} 
When the algorithm above identifies a combination that satisfies condition (\ref{eq:condition}), the symmetry analysis for the selected set of {\it relative terms} of the PES is completed having determined: i) the $n_0$ terms that are null by symmetry, ii) the minimal set of $n_\textup{ec}$ terms to be explicitly computed via the EGH scheme that allow all other $n_\textup{s}$ terms to be obtained from Eq. (\ref{eq:solution}). The algorithm then moves to the next set of {\it relative terms} until all terms of the PES in Eq. (\ref{eq:pes}) have been analysed.

Upon completion of this symmetry analysis for all terms of the PES in Eq. (\ref{eq:pes}), we are ready to determine what nuclear configurations can be skipped in the EGH procedure sketched in Figure \ref{fig:egh}. At Step 1 of the EGH procedure, for each selected normal mode $i=1,\dots,M$, the algorithm checks if both the corresponding 1M terms $\eta_{iii}$ and $\eta_{iiii}$ can be obtained as solutions of the linear systems of equations discussed above. If and only if that is the case, then the calculations at the two nuclear configurations obtained by displacing the atoms along the $i$-th normal mode (blue circles in Figure \ref{fig:egh}) can be skipped as $\eta_{iii}$ and $\eta_{iiii}$ do not need to be explicitly computed via the EGH scheme through Eqs. (\ref{eq:scheme3step11}-\ref{eq:scheme3step12}). At Step 2 of the EGH procedure, for each selected pair of normal modes $ i<j =1,\dots, M$, the algorithm checks if all the corresponding 2M terms $\eta_{ijj}$, $\eta_{iij}$, $\eta_{iiij}$, $\eta_{ijjj}$ and $\eta_{iijj}$ can be obtained as solutions of the linear systems of equations discussed above. If and only if that is the case, then the calculations at the two nuclear configurations obtained by simultaneously displacing the atoms along the $i$-th and $j$-th normal modes (green circles in Figure \ref{fig:egh}) can be skipped as the 2M terms do not need to be explicitly computed via the EGH scheme through Eqs. (\ref{eq:scheme3step21}-\ref{eq:scheme3step23}).

\section{Results and Discussion} 

In this section, we present examples on the application to molecules and solids of the symmetry analysis described in Section \ref{sec:form} to reduce the number of anharmonic terms of the PES to be explicitly computed via the numerical EGH finite-difference scheme. We start by providing examples on the use of the symmetry relations introduced in Section \ref{subsec:symm} and then we present the computational gains obtained from their exploitation in actual calculations.

\subsection{Examples on the Use of the Symmetry Relations}

We discuss a couple of explicit examples on how the symmetry relations obtained with Eq. (\ref{eq:general_symm_rel}) can be exploited so as to make the algorithm described in Section \ref{subsec:use} clearer by exemplification. 

We start from the methane molecule, whose normal modes have already been characterized by symmetry in Table \ref{tab:methane}. Let us consider quartic terms $Q_iQ_jQ_kQ_l$ where the first two modes ($Q_i$ and $Q_j$) belong to the first {\it set} (i.e. the first occurrence of the 3D irrep $F_2$) and the  last two modes ($Q_k$ and $Q_l$) belong to the second {\it set} (i.e. the first occurrence of the 2D irrep $E$). There is a total of $n_\textup{rt} =18$ potentially distinct terms of this kind that form a group of {\it relative terms}: six 2M terms, nine 3M terms, and three 4M terms. Application of the invariance condition (\ref{eq:general_symm_rel}) to each of the 18 terms leads to the linear system of equations (\ref{eq:system2}). The corresponding 18$\times$18 $\bf{C}^\prime$ matrix is reported in Figure \ref{fig:mat} A$_1$. Here a compact notation is used to label each relative term by use of the index of the row for each mode, where the indices of the modes belonging to the second {\it set} are primed. For instance, the term $Q_1Q_1Q_4Q_4$ (a 2M term) is simply labeled $111^\prime1^\prime$ because mode 1 corresponds to the first row of the irrep $F_2$ and mode 4 corresponds to the first row of the irrep $E$; accordingly, the term $Q_1Q_3Q_4Q_5$ (a 4M term) is labeled $131^\prime2^\prime$, and so on. Inspection of Figure \ref{fig:mat} A$_1$ reveals that 10 out of 18 terms are null by symmetry (i.e. $n_0 =10$) and therefore must not be explicitly computed. By removing the $n_0$ terms from the linear system we are left with an 8$\times$8 matrix shown in Figure \ref{fig:mat} A$_2$. We now look for the minimal set of terms to be explicitly computed, which makes this system solvable via the algorithm described in Section \ref{subsec:use}. In this case, it turns out that by explicitly computing just two terms, namely $111^\prime1^\prime$ and $112^\prime2^\prime$ (first and third columns in Figure \ref{fig:mat} A$_2$), and by removing the first two rows in Figure \ref{fig:mat} A$_2$, the resulting reduced matrix $\bf{C}^\prime_{\bf r}$ has a non null determinant, as shown in Figure \ref{fig:mat} A$_3$, and therefore allows the system to be solved. To summarize, the symmetry analysis of these {\it relative terms} tells us that out of the total of 18 terms, 10 are null by symmetry and that six of the remaining eight non-null ones can be derived by exploitation of symmetry relations by explicitly computing only two of them via the EGH scheme.

\begin{widetext}
\begin{center}
\begin{figure}[h!!]
\centering
\includegraphics[width=14cm]{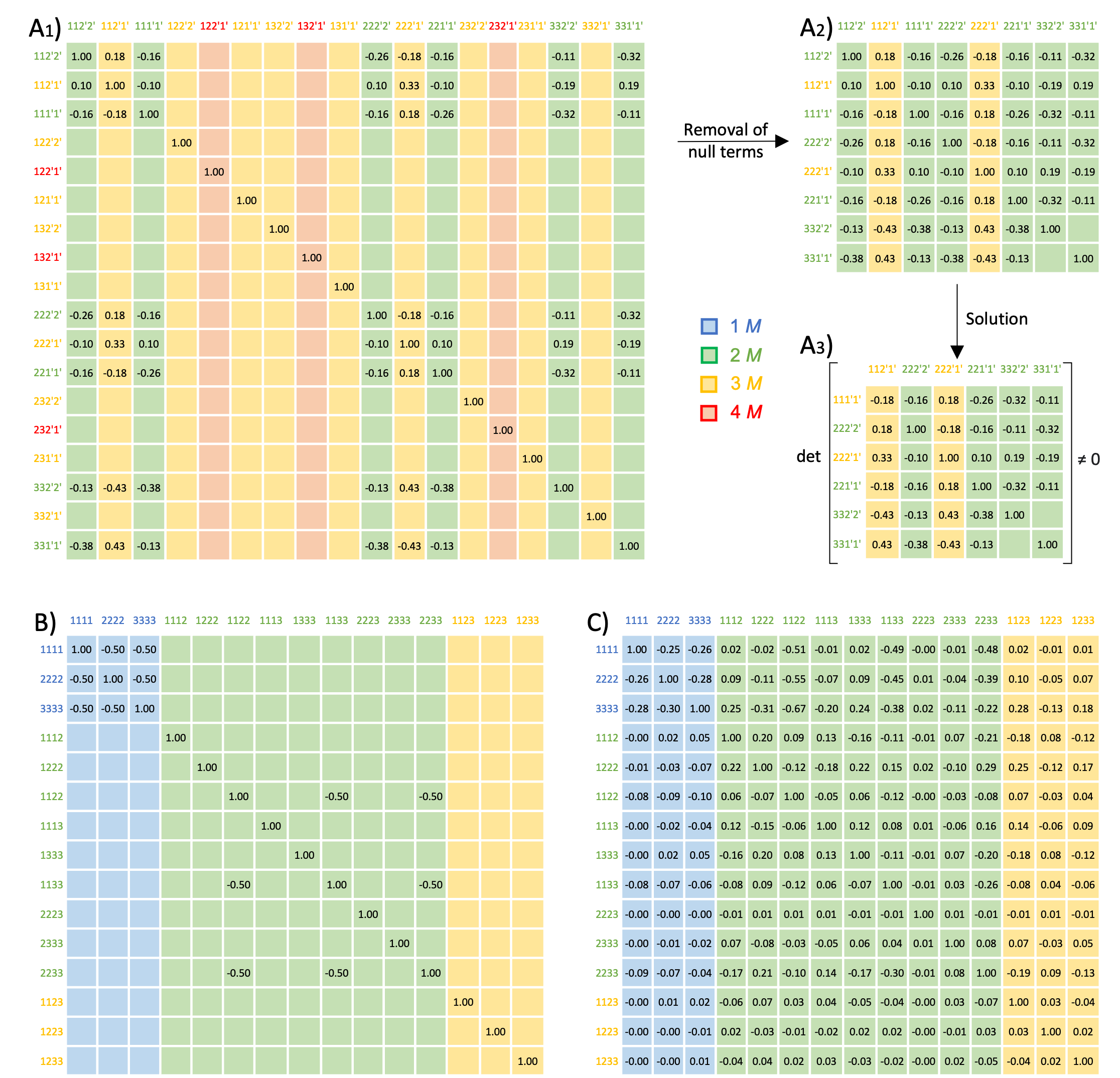}
\caption{Examples of linear systems of equations of the form (\ref{eq:system2}), obtained from the invariance condition (\ref{eq:general_symm_rel}), for selected molecules and selected anharmonic terms of the PES. A$_1$) Linear system for quartic terms of the anharmonic PES of methane involving two modes of the first {\it set} (3D) and two modes of the second {\it set} (2D), see Table \ref{tab:methane}. A$_2$) Same as in A$_1$) but with null terms being removed. A$_3$) Corresponding $\bf{C}^\prime_{\bf r}$ matrix, which satisfies condition (\ref{eq:condition}). B) Linear system for quartic terms of the anharmonic PES of methane involving only modes of the first {\it set} (3D). C) Linear system for quartic terms of the anharmonic PES of tetrahedrane involving only modes of the third {\it set} (3D). The following color scheme is used: blue for 1M, green for 2M, yellow for 3M and red for 4M terms, respectively. Empty boxes correspond to null elements in the matrices.}
\label{fig:mat}
\end{figure}
\end{center}
\end{widetext}

For a second example, let us still work with methane. Let us now consider quartic terms $Q_iQ_jQ_kQ_l$ where all four modes belong to the first {\it set} (i.e. the first occurrence of the 3D irrep $F_2$). In this case, there is a total of $n_\textup{rt} =15$ potentially distinct terms of this kind that form a group of {\it relative terms}: three 1M terms, nine 2M terms, and three 3M terms. Application of the invariance condition (\ref{eq:general_symm_rel}) to each of the 15 terms leads to the linear system of equations (\ref{eq:system2}). The corresponding 15$\times$15 $\bf{C}^\prime$ matrix is reported in Figure \ref{fig:mat} B. The same compact notation introduced above is used to label each relative term. Inspection of Figure \ref{fig:mat} B reveals the peculiar block-diagonal form of the matrix where 1M terms are symmetry related only to other 1M terms, 2M terms only to 2M terms, while all 3M terms are null by symmetry in this case. Each block can thus be analysed independently. For the three 1M terms, it turns out that it is enough to explicitly compute the first one (namely 1111, that is $Q_1Q_1Q_1Q_1$) to make the system solvable and thus to obtain the other two (2222 and 3333, that is $Q_2Q_2Q_2Q_2$ and $Q_3Q_3Q_3Q_3$). For the nine 2M terms, six of them are null by symmetry, and also in this case it turns out that it is enough to explicitly compute the first non-null one (namely 1122, that is $Q_1Q_1Q_2Q_2$) to make the system solvable and obtain the other two non-null ones (1133 and 2233, that is $Q_1Q_1Q_3Q_3$ and $Q_2Q_2Q_3Q_3$). To summarize, the symmetry analysis of these {\it relative terms} tells us that out of the total of 15 terms, 9 are null by symmetry and that four of the remaining six non-null ones can be derived by exploitation of symmetry relations by explicitly computing only two of them via the EGH scheme.

As a last example, we consider the tetrahedrane molecule, C$_4$H$_4$, also belonging to the T$_\textup{d}$ point symmetry group. Let us consider quartic terms $Q_iQ_jQ_kQ_l$ where all four modes belong to the third {\it set} (i.e. the first occurrence of the 3D irrep $F_1$). As in the previous case, there is a total of $n_\textup{rt} =15$ potentially distinct terms of this kind that form a group of {\it relative terms}: three 1M terms, nine 2M terms, and three 3M terms. Application of the invariance condition (\ref{eq:general_symm_rel}) to each of the 15 terms leads to the linear system of equations (\ref{eq:system2}). The corresponding 15$\times$15 $\bf{C}^\prime$ matrix is reported in Figure \ref{fig:mat} C. Inspection of Figure \ref{fig:mat} C reveals another interesting structure, where no terms are found to be null by symmetry and where all symmetry relations link 1M, 2M and 3M terms together. As complex the symmetry relations may look in this case, this system can be solved by explicitly computing a surprisingly low number of terms. Indeed, by computing via the EGH scheme just the first two terms (namely 1111 and 2222, that is two 1M terms), the system becomes solvable and all other 13 terms can be derived.

The few selected examples discussed above show the effectiveness of an {\it a priori} symmetry analysis in reducing the number of anharmonic terms of the PES that need to be explicitly computed, with associated computational gains to be documented in the following section.

\subsection{Computational Gain}

We have implemented in a developmental version of the \textsc{Crystal23} program the group theoretical approach described in Section \ref{sec:form} to simplify the numerical calculation of cubic and quartic anharmonic terms of the PES. In this Section, we illustrate its effectiveness in reducing the number of atomically displaced nuclear configurations $N_\textup{conf}$ at which the energy and forces must be computed. Four molecular systems are considered: methane, CH$_4$ (belonging to the T$_\textup{d}$ point symmetry group, with 9 normal modes), tetrahedrane, C$_4$H$_4$ (belonging to the T$_\textup{d}$ point symmetry group, with 18 normal modes), cyclo-exasulfur, S$_6$ (belonging to the D$_\textup{3d}$ point symmetry group, with 12 normal modes), and cubane, C$_8$H$_8$ (belonging to the O$_\textup{h}$ point symmetry group, with 42 normal modes). Two 3D crystalline solids are also considered: Magnesium oxide, MgO, as described by a conventional cubic cell (belonging to the Fm$\overline{3}$m cubic space group, with 21 normal modes), and a Zinc-imidazolate framework, namely ZIF-8 (belonging to the I$\overline{4}$3m cubic space group). In the latter case, given that ZIF-8 has 138 atoms per primitive cell and thus a total of 411 normal modes, a sub-set of just 12 modes has been selected for the anharmonic analysis, corresponding to the highest frequency ones. Figure \ref{fig:gain} shows the atomic structure of the six selected systems. For each system, the figure also shows bar plots reporting the total number of nuclear configurations $N_\textup{conf}$ that need to be explicitly explored within the EGH finite-difference scheme when symmetry is not exploited and when symmetry is exploited according to the group theoretical approach presented here. 

In the case of methane, CH$_4$, there are $M=9$ normal modes. Within a 2M4T representation of the PES, there are a total of 198 cubic and quartic anharmonic force constants to be computed. Without the exploitation of symmetry, 90 atomically displaced nuclear configurations should be explored, which are reduced to 30 by symmetry exploitation as described in Section \ref{sec:form}, with a computational gain that amounts to a factor of 3. For tetrahedrane, C$_4$H$_4$, there are $M=18$ normal modes. Within a 2M4T representation of the PES, there are a total of 801 cubic and quartic anharmonic force constants to be computed. Without the exploitation of symmetry, 342 distinct atomically displaced nuclear configurations should be explored, which are reduced to 110 by symmetry exploitation, with a computational gain that amounts to a factor of 3.1. The cyclo-exasulfur, S$_6$, molecule has $M=12$ normal modes. Within a 2M4T representation of the PES, there are a total of 354 cubic and quartic anharmonic force constants to be computed. Without the exploitation of symmetry, 156 distinct atomically displaced nuclear configurations should be explored, which are reduced to 96 by symmetry exploitation, with a computational gain that amounts to a factor of 1.6. In the case of cubane, C$_8$H$_8$, there are $M=42$ normal modes. Within a 2M4T representation of the PES, there are a total of 4389 cubic and quartic anharmonic force constants to be computed. Without the exploitation of symmetry, 1806 distinct atomically displaced nuclear configurations should be explored, which are reduced to 566 by symmetry exploitation, with a computational gain that amounts to a factor of 3.2. For the MgO crystal, described by its cubic conventional cell with 8 atoms, there are $M=21$ normal modes. Within a 2M4T representation of the PES, there are a total of 1092 cubic and quartic anharmonic force constants to be computed. Without the exploitation of symmetry, 462 distinct atomically displaced nuclear configurations should be explored, which are reduced to 112 by symmetry exploitation, with a computational gain that amounts to a factor of 4.1. Finally, for the ZIF-8 crystal, by considering only a subset of 12 normal modes, there are a total of 354 cubic and quartic anharmonic force constants to be computed within a 2M4T representation of the PES. Without the exploitation of symmetry, 156 distinct atomically displaced nuclear configurations should be explored, which are reduced to 48 by symmetry exploitation, with a computational gain that amounts to a factor of 3.2.

In conclusion, the effectiveness of the approach to reduce the number of atomically displaced nuclear configurations to be explicitly explored depends on two factors: i) the order of the point group of the system: clearly, the richer the point symmetry, the higher the speedup; ii) the occurrence of high-dimensional (2D or 3D) irreps in the manifold of normal modes.

\begin{widetext}
\begin{center}
\begin{figure}[h!!]
\centering
\includegraphics[width=17cm]{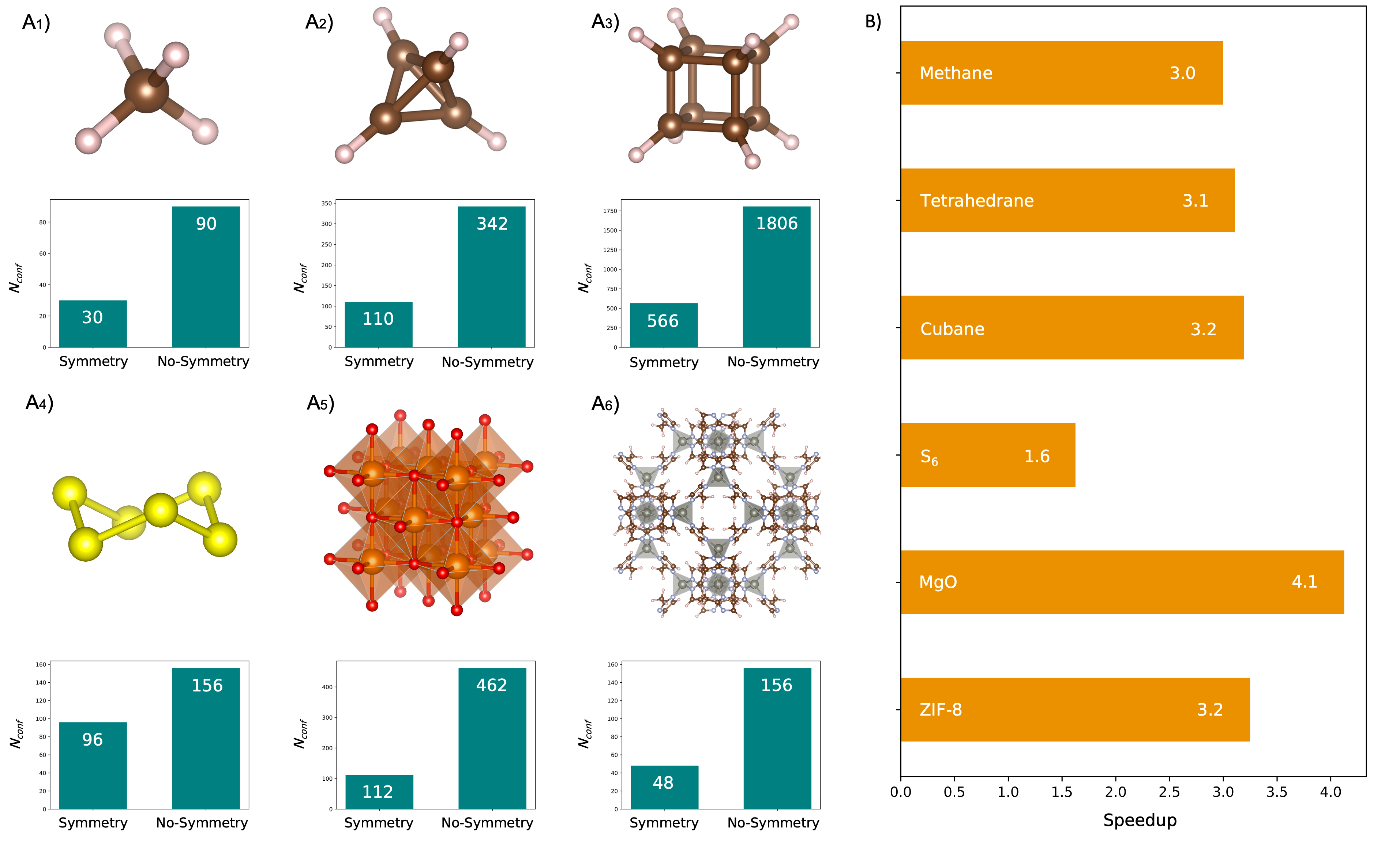}
\caption{A) For each of the six test systems, the atomic structure is shown and a bar plot is presented, which reports the total number of nuclear configurations $N_\textup{conf}$ that need to be explicitly explored in building the 2M4T anharmonic PES with the EGH finite-difference scheme when symmetry is not exploited and when symmetry is exploited according to the group theoretical approach presented here. B) Overall computational gain factor due to the symmetry exploitation approach discussed in Section \ref{sec:form}.}
\label{fig:gain}
\end{figure}
\end{center}
\end{widetext}

\section{Conclusions}

A group theoretical approach based on Wigner's theorem has been formally illustrated to determine symmetry relations among anharmonic terms of the potential energy surface (PES) of a quantum-mechanical system. An algorithm has been devised to take full advantage of such symmetry relations to reduce the number of atomically displaced nuclear configurations at which energy and forces must be computed to build a 2M4T representation of the anharmonic PES, with the EGH finite-difference scheme. The algorithm has been implemented in a developmental version of the \textsc{Crystal23} software program, and tested on six high symmetry systems (four molecules and two 3D crystals). The results clearly demonstrate the consistent computational gain provided by such approach for highly symmetric systems. The highest speedup (factor of 4.1) has been obtained for a cubic crystal of magnesium oxide.

\acknowledgements

A.E. wishes to dedicate this paper to the memory of Prof. Claudio Zicovich-Wilson, for introducing him to the wonders of group theory in quantum chemistry. A.E. gratefully acknowledges insightful discussions with Prof. Michel R\'{e}rat on the application of group theory to anharmonic terms of the potential energy surface. A.E. and J.M. thank the University of Torino and the Compagnia di San Paolo for funding (CSTO169372).

%\bibliographystyle{achemso}
%\bibliography{DatabaseBIBLIO,DatabaseBIBLIO_anh1}

\providecommand{\latin}[1]{#1}
\makeatletter
\providecommand{\doi}
  {\begingroup\let\do\@makeother\dospecials
  \catcode`\{=1 \catcode`\}=2 \doi@aux}
\providecommand{\doi@aux}[1]{\endgroup\texttt{#1}}
\makeatother
\providecommand*\mcitethebibliography{\thebibliography}
\csname @ifundefined\endcsname{endmcitethebibliography}
  {\let\endmcitethebibliography\endthebibliography}{}

\end{document}